\documentclass[aip,apl,noshowpacs,amsmath,amssymb,twocolumn,superscriptaddress,raggedbottom,floatfix]{revtex4}
\usepackage{graphicx}
\usepackage{color}

\begin{document}

\title{Field-effect diode based on electron-induced Mott transition in NdNiO$_3$}

\author{W.~L.~Lim}
\affiliation{Department of Physics, Emory University, Atlanta, GA 30322, USA}

\author{E.~J.~Moon}
\affiliation{Physics Department, University of Arkansas, Fayetteville, AR 72701, USA}

\author{J.~W.~Freeland}
\affiliation{Advanced Photon Source, Argonne National Laboratory, Argonne, IL 60439, USA}

\author{D.~J.~Meyers}
\affiliation{Physics Department, University of Arkansas, Fayetteville, AR 72701, USA}

\author{M.~Kareev}
\affiliation{Physics Department, University of Arkansas, Fayetteville, AR 72701, USA}

\author{J.~Chakhalian}
\affiliation{Physics Department, University of Arkansas, Fayetteville, AR 72701, USA}

\author{S.~Urazhdin}
\email[]{Sergei.Urazhdin@emory.edu}
\affiliation{Department of Physics, Emory University, Atlanta, GA 30322, USA}

\begin{abstract}
We studied an electron-induced metal-insulator transition in a two-terminal device based on oxide NdNiO$_3$. In our device, the NdNiO$_3$ is electrostatically doped by the voltage applied between the terminals, resulting in an asymmetric conductivity with respect to the bias polarity. The asymmetry is temperature-dependent and is most significant near the metal-insulator transition. The \textit{I-V} characteristics exhibit a strong dependence both on the thermal history and the history of the applied voltage bias. Our two-terminal device represents a simple and efficient route for studies of the effect of electron doping on the metal-insulator transition.
\end{abstract}


\maketitle


The ability to electronically control the resistance of materials is the fundamental mechanism underlying the operation of modern semiconductor devices. In addition, a number of advanced materials and structures that exhibit controllable resistance switching phenomena are currently being explored as a basis for the future devices that will combine robust processibility, scalability, energy efficiency, and high speed.  In strongly correlated oxide materials exhibiting a metal-insulator transition (MIT), phase transitions between states with different transport and magnetic properties \cite{Yang2011,Yang2012,ahn_electrostatic_2006} provide a fundamental mechanism for the operation of electronic devices with advanced functionalities. Such devices can be controlled by electric field,\cite{ahn_electric_2003, chen_resistance_2010, ruzmetov_three-terminal_2010} optical excitation,\cite{Lysenko_S_2007} or a combination of thermal and electronic effects.\cite{cao_thermoelectric_2009, yajima_2012}  The electric field can not only modulate the charge density, but also shift the phase transition temperature, thus enhancing the direct effects of the electric field on the conductivity. The effects of electrical gating on the MIT have been demonstrated in La$_{0.8}$Ca$_{0.2}$MnO$_3$ films backgated through the substrate, and in NdNiO$_3$ (NNO) films gated through ionic liquids.~\cite{koetz_principles_2000, dhoot_beyond_2006, misra_electric_2007, ueno_electrostatic_2010, Asanuma_2010} The electric field-induced variations of MIT temperature $T_{MI}$ resulted in a strong variation of the phase-dependent charge density.~\cite{scherwitzl_electricfield_2010} These experiments provided a significant fundamental insight into the interplay between the charge carrier density and phase transitions in correlated oxides.

Studies of electric field effects have so far focused on three-terminal field-effect transistor (FET) structures  incorporating a gate electrode separated from the oxide film by an electrically insulating gate dielectric.  However, these studies have been constrained by the difficulty of incorporating complex oxides into the standard three-terminal FET structure with a separate gate electrode. Here, we demonstrate a simple unipolar \emph{two-terminal} device, a field-effect diode (FED), that enables studies of field effects in correlated oxides. The device is more straightforward than a FET both in terms of the fabrication and measurements. In our device geometry, the electric field is determined by the voltage applied between the two terminals. Thus, by measuring the dependence of the device conductivity on the applied voltage, one can determine the effects of the electric field. Here, we exemplify the device operation with the unipolar switching of a FED based on the Mott insulator NNO. We demonstrate unique electronic transport characteristics that appear at temperatures close to $T_{MI}$, elucidating the effects of electric field on the MIT.

The field-effect diode is based on a 20-unit-cell (7.8nm) thick NNO film grown by the pulsed laser deposition technique. Fabrication of advanced functional devices requires high quality materials with excellent epitaxial, stoichiometric, and morphological characteristics.  In particular, epitaxial layer-by-layer growth of ultrathin NNO films on STO is critical for creating  tensile strain over the entire NNO film, leading to a robust MIT.~\cite{liu_strain-mediated_2010} To achieve high-quality epitaxial  growth, an atomically flat TiO$_2$-terminated STO(001) single crystal was prepared by a wet-etch procedure designed to minimize the surface and near-surface electronic defects.~\cite{kareev_atomic_2008}  The deposition was monitored by in-situ high pressure reflection high energy electron diffraction (HP-RHEED) capable of operation at oxygen pressures of up to 400 mTorr. After the deposition, the NNO films were annealed in one atmosphere of ultra-pure oxygen to minimize the oxygen deficiency that adversely affects the electronic properties.~\cite{misha_jap_2011}

\begin{figure}[]
\includegraphics[width=\columnwidth]{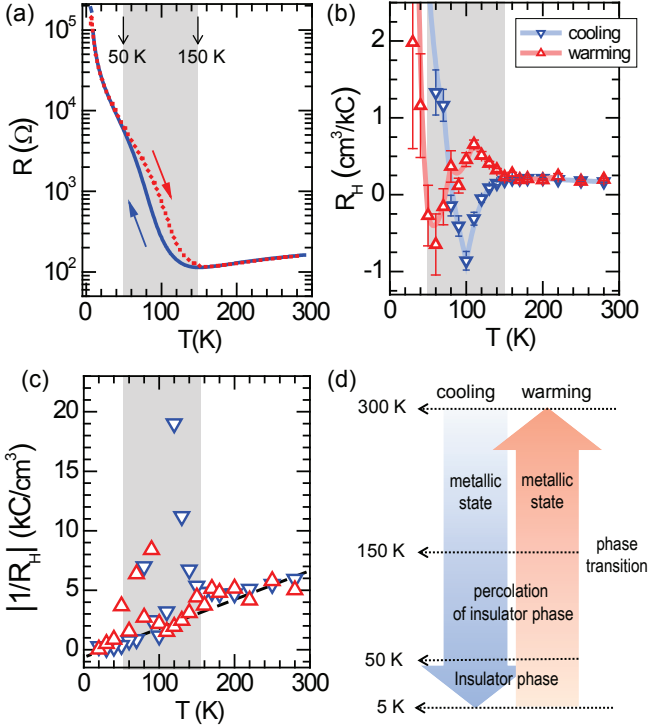}
\caption{\label{fig:fig1} (Color online) (\emph{a}) Temperature dependence of  resistance for a $1$cm$\times1$cm extended NNO/STO film demonstrating a hysteretic MIT at 150~K. The transport characteristics were measured in the \textit{Van der Pauw} geometry. The shaded area between 50~K and 150~K indicates the region of percolation of insulating phase in the metallic state. (\emph{b}) Hall coefficient and (\emph{c}) Hall carrier density of the  film. Note that in the insulating phase at $T<50$~K, the uncertainty of measurement is very large due to the high resistance. Dashed line in (\emph{c}) indicates a quasi-linear temperature dependence of the carrier concentration. (\emph{d}) Diagram showing the relationship between temperature and the phase transitions in NNO upon cooling and warming.}
\end{figure}

To understand the MIT-related properties of NNO that are essential for the operation of FED device, we characterized the as-deposited unpatterned NNO film by electronic transport measurements (Fig.~\ref{fig:fig1}(\emph{a-c})). The resistance [Fig.~\ref{fig:fig1}(\emph{a})] and the Hall coefficient $R_H$ [Fig.~\ref{fig:fig1}(\emph{b})] were measured at both increasing and decreasing $T$. The resistance of the film increased with increasing temperature below $150$~K, and increased above it, consistent with the usual MIT behavior in high-quality NNO films. The thermal hysteresis of resistance in the temperature range between $50$~K and $150$~K (highlighted by a grey region in Figs.~\ref{fig:fig1} (\emph{a-c}) ) is consistent with the first-order nature of the metal-insulator transition. We identify the transition temperature $T_{MI}=150$~K by the sign change of the slope of the resistance versus temperature curve.

The Hall coefficient $R_H$ is positive away from the transition point, indicating that the charge carriers are predominantly $p$-type.~\cite{scherwitzl_electricfield_2010}  Strikingly, the sign of $R_H$ reverses just below the MIT. A similar sign change of the Hall coefficient was observed in the vicinity of MIT in VO$_2$.~\cite{vo2} Sign changes of $R_H$ have been also observed in the charge density wave and the fluctuation regimes of the high temperature superconductors, and were attributed to the opening of a pseudogap.~\cite{hns_lee_1970, naito_1982, leboeuf_2007}
The sign change in NNO can be similarly attributed to the Fermi surface reconstruction associated with the phase transition, leading to the opening of the Mott-Hubbard gap. The emergence of the gap combined with the complex multiband electronic structure of NNO, can result in coexistence of multiple carrier types near $T_{MI}$.~\cite{moon_2012}

The inverse of Hall coefficient $R_H^{-1}(T)$ provides direct information about the carrier concentration.  As shown in Fig.~\ref{fig:fig1} (\emph{c}), two regimes can be identified based on the temperature dependence of $R_H^{-1}(T)$. The first regime at both low and high temperature away from the metal-insulator transition is characterized by a quasi-linear dependence (dashed line in Fig.~\ref{fig:fig1}(\emph{c})), as also observed in other nickelates.~\cite{moon_2012} In contrast, $R_H^{-1}(T)$ exhibits strong nonlinear variations in the temperature range between $50$~K and $150$~K (grey area in Figs.~\ref{fig:fig1} (\emph{c})). Such strong variations can be attributed to the significant changes of carrier densities and dominant types in the region of the percolation of insulating phase in the metallic state. The evolution of phases with decreasing and increasing temperatures, deduced from the electronic measurements, is schematically summarized in the diagram in Fig.~\ref{fig:fig1} (\emph{d}).

\begin{figure}[t]
\includegraphics[width=\columnwidth]{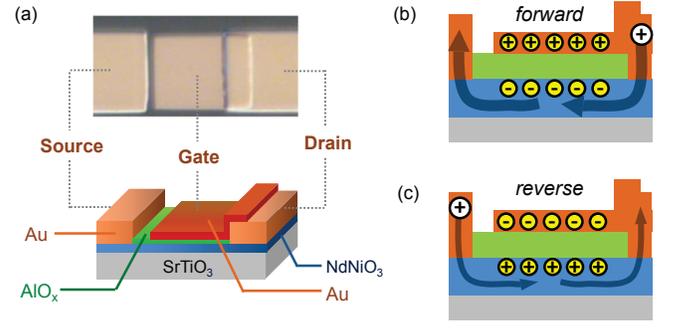}
\caption{\label{fig:fig2} (Color online) (\emph{a}) Schematic and an optical micrograph of the NNO/STO-based field-effect diode structure. (\emph{b}) Forward-bias regime of the device: electron accumulation is induced in the NNO channel by a positive drain-source voltage $V_{DS}$. (\emph{c}) Reverse-bias regime at $V_{DS}<0$.}
\end{figure}

\begin{figure}[t]
\includegraphics[width=\columnwidth]{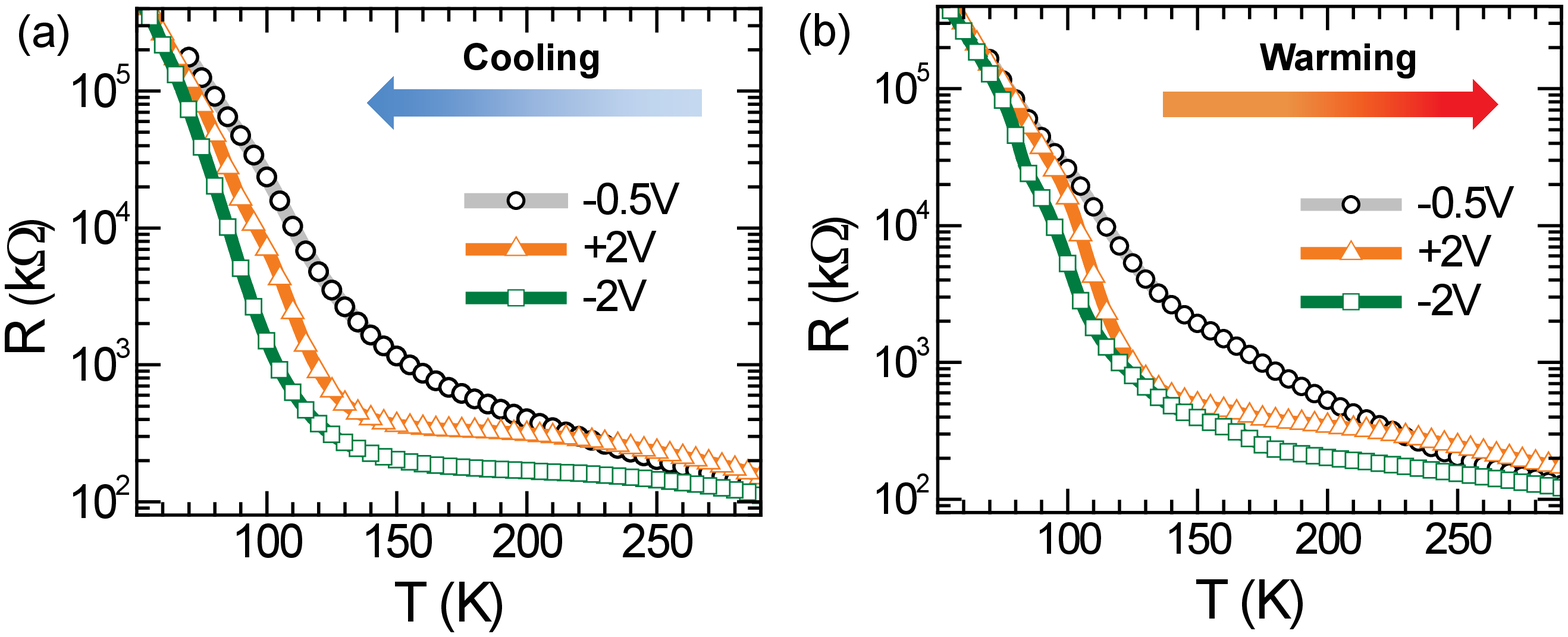}
\caption{\label{fig:fig3} (Color online) Temperature dependence of the device resistance upon cooling (\emph{a}) and warming (\emph{b}).}
\end{figure}

The schematic of the field-effect diode (FED) and its operation principle are shown in Fig.~\ref{fig:fig2}. The device electrodes were defined on the NNO film by a combination of \textit{e}-beam lithography and sputtering of a Ni(5)Au(150) bilayer, where thicknesses are in nm. The Ni layer improved the adhesion of the contact and was also essential for making an ohmic contact to NNO, as was verified by comparing two-probe and four-probe measurements on a separate NNO test film. We subsequently deposited a $10\mu$m$\times10\mu$m AlO$_x$(6) layer serving as a gate insulator. In the final step, the gate insulator was covered by a Ni(5)Au(50) bilayer gate that was connected to one of the device electrodes. In the following, we will refer to this electrode as the drain, and to the other electrode as the source.

When a voltage $V_{DS}$ is applied between the drain and the source, it produces an electric field between the gate and the NNO channel. This field induces electron accumulation in the NNO channel in forward-bias, and electron depletion  in reverse-bias, as illustrated in Figs.~\ref{fig:fig2} (\emph{b}) and (\emph{c}). As a consequence, the electronic properties of the FED depend on the magnitude and the polarity of the applied voltage. We emphasize that our device does not rely on gating with ionic liquids, avoiding chemical deterioration of the interface,~\cite{Yang2012} and enabling us to determine the effects of varying electric field at a fixed temperature. We will show below that this ability provides an additional insight into the effects of electric field on the  MIT.

The dependence of resistance on temperature is qualitatively similar to the extended film, but instead of a minimum, the resistance exhibits an inflection at $T_{MI}$ with a gradual decrease at larger $T$ (Fig.~\ref{fig:fig3}). The disappearance of the resistance minimum is likely caused by the additional charge scattering at the interface with the AlO$_x$ gate insulator. We note that the resistance varies by three orders of magnitude between room temperature and $5$~K, consistent with the behaviors of unpatterned films (Fig.~\ref{fig:fig1}(a)). Therefore, we conclude that the electric contacts to the device are ohmic and give a negligible contribution to the measured device resistance.

The resistance depends not only on temperature, but also on the drain-source voltage $V_{DS}$, as illustrated for three representative values $V_{DS}=+2$~V, $-0.5$~V, and $-2$~V as shown in Fig.~\ref{fig:fig3}. The resistance exhibits the strongest relative dependence on $V_{DS}$ in the vicinity of $T_{MI}$. For instance, when the temperature is decreased to $T=106$~K (Fig.~\ref{fig:fig3} (a)), the resistance at $V_{DS}=-2$~V is $17$ times smaller than at $V_{DS}=-0.5$~V. The effects of $V_{DS}$ are also different for decreasing temperature [Fig.~\ref{fig:fig3}(\emph{a})] as compared to increasing temperature [Fig.~\ref{fig:fig3}(\emph{b})]. For instance, when increasing the temperature to $133$~K, the resistance is the same at $V_{DS}=+2$~V and $V_{DS}=-2$~V. Meanwhile, when decreasing the temperature to $133$~K, the resistance at $+2$~V is $1.9$ times larger than at $-2$~V. The strong hysteretic response of the device to electric field near MIT indicates the influence of the electric field on the first-order phase transition.

\begin{figure}
\includegraphics[width=\columnwidth]{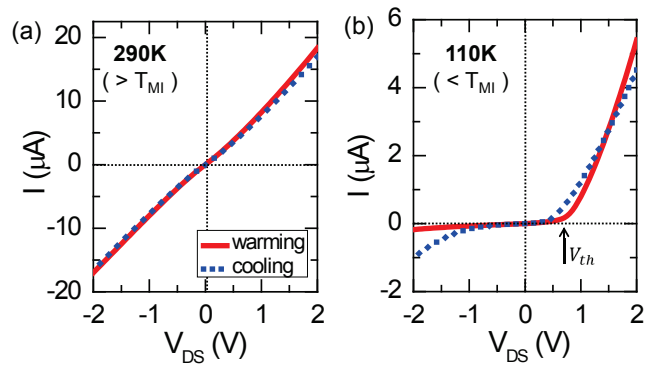}
\caption{\label{fig:fig4} (Color online) \textit{I-V} curves obtained after increasing $T$ from $6$~K (solid red), and decreasing $T$ from $295$~K (dashed blue), at (\emph{a}) $290$~K ($> T_{MI}$) and (\emph{b}) $110$~K ($< T_{MI}$). Conductance threshold voltage $V_{th}$ is marked by an arrow.}
\end{figure}

Figure~\ref{fig:fig4} shows \textit{I-V} curves for two representative temperatures $T = 290$~K and $110$~K. The linear \textit{I-V} characteristics at $T=290$~K are consistent with the metallic phase of NNO at high temperatures [Fig.~\ref{fig:fig4}(\emph{a})], and only a small effect of the electric field on conductance. However, at $T = 110$~K~$<T_{MI}$, the \textit{I-V} characteristics become significantly nonlinear and asymmetric [Fig.~\ref{fig:fig4}(\emph{b})]. At small bias voltages below a threshold value $V_{th}=0.7$~V, the device exhibits insulating behaviors characterized by a small conductance $I/V<0.083$~$\mu$S. The increase of conductance at larger bias is more significant at $V_{DS}>0$ than at $V_{DS}<0$, resulting in asymmetric nonlinear \textit{I-V} characteristics similar to the bipolar diodes. Thus, the device behaves like a $resistor$ above $T_{MI}$ and becomes a $diode$ below $T_{MI}$. In contrast to semiconductor \textit{p-n} diodes, our device incorporates a single channel material (NNO) and thus represents a \textit{unipolar field-effect} diode.

The observed dependence of the conductivity on the magnitude and sign of bias are indicative of the effects of electron/hole doping in NNO on the MIT. In particular, the increases of conductivity at both positive and negative bias are consistent with the breakdown of the critical Coulomb interaction between electrons, resulting in the transition to a metallic state. Moreover, a smaller  enhancement associated with hole doping at negative bias suggests the possibility of insulating state that accommodates extra holes e.g. by formation of a charge density wave without a breakdown of electron correlations~\cite{kim}.

The asymmetry of \textit{I-V} characteristics is stronger for data obtained for increasing temperature (solid red curve in Fig.~\ref{fig:fig4}(\emph{b})) than for decreasing temperature (dash blue curve in Fig.~\ref{fig:fig4}(\emph{b})). The rectifying effect is most significant in the vicinity of MIT. The on/off ratio, defined as $I(V_{DS}=+2~V)/I(V_{DS}=-2~V)$, is about $300$ at $T=110$~K when the device is warmed from $6$~K. Several important conclusions can be made based on these observations. First, the thermal hysteresis is consistent with the first-order nature of the MIT. Second, since the diode-like behaviors are most pronounced in the vicinity of MIT, we can conclude that these behaviors are associated with the effect of electric field on the phase transition. Third, since the conductance of the device is higher at positive bias than at negative bias, we conclude that the insulating state is suppressed, and the metallic state is enhanced by the electron doping. Fourth, since the electric field effects are stronger upon warming than upon cooling, we conclude that these effects are most significant when the material is in the insulating state percolated by the metallic phase.

\begin{figure}
\includegraphics[width=\columnwidth]{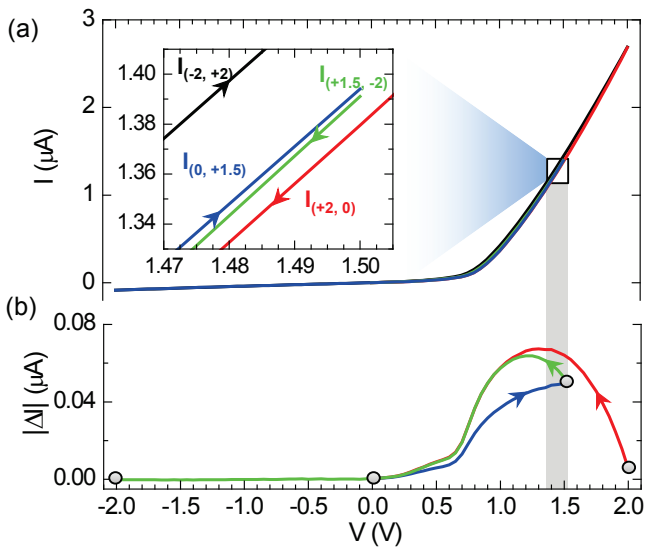}
\caption{\label{fig:fig5} (Color online) (\emph{a})
Hysteretic \textit{I-V} characteristics at $T=100$~K, obtained by sequential scanning of the bias voltage from  $V_{DS}=-2$~V to $2$~V, $0$~V, $1.5$~V, and finally back to $-2$~V. The current measured while scanning between the bias values $u$ and $v$ is labeled $I_{(u, v)}$.  Inset shows \textit{I-V} data close to $V_{DS}=1.5$~V. (\emph{b}) Relative differences $|\Delta I|$ between the initial scan and each of the subsequent scans.}
\end{figure}

The rectifying behaviors of FED depend not only on the thermal history, but also on the history of the applied bias, as illustrated by the sequence of bias scans in Fig.~\ref{fig:fig5}. After initially setting $V_{DS}$ to $-2$~V, the bias was  scanned first to $V_{DS}=+2$~V, then decreased to $0$~V, then increased again to $1.5$~V, and finally decreased to $-2$~V. In this sequence, the range between $V_{DS}=0$~V and $V_{DS}=+1.5$~V was scanned four times and each scan yielded a different conductance at a given bias. To highlight the differences, $|\Delta I|$, the initial $I_{(-2, +2)}$ scan from $-2$ to $+2$~V was subtracted from the subsequent scans, $I_{(i, j)}$. Comparing only the two upward scans, $I_{(-2, +2)}$ and $I_{(0, +1.5)}$, the positive-bias conductivity was larger when starting from a larger negative bias due to the electron-doping rather than hole-doping. In the two downward scans, $I_{(+2, 0)}$ and $I_{(+1.5, -2)}$, the conductivity at +1.5~V was lower than that at the same bias in the downward scan $I_{(+2, 0)}$, but merged below 0.7~V. The \textit{I-V} curves at $V_{DS}<0$~V were independent of the bias history. These results are consistent with the above conclusion that electric field influences the geometry of the phase percolation in NNO, and that the effects are most significant in the electron enhancement regime.

To summarize, we have demonstrated significant effects of electrostatic doping on the metal-insulator transition in a two-terminal solid-state device based on a thin NdNiO$_3$ film. Our device exhibits diode-like rectifying behaviors that depend on the thermal history, consistent with the first-order nature of the metal-insulator transition. In addition, the rectifying behaviors depend on the history of the applied bias, indicating direct effects of electrostatic doping on the phase transition. Our device represents a simple and efficient route for studies of field effects in thin complex oxide films that are generally not amenable to solid-state engineering.

The research at the University of Arkansas was supported by ONR grant (No. 10-001-SA1002031), and partially by  NSF grant (No. DMR-0747808) and  DOD-ARO grant (No. W911NF-11-1-0200).



\end{document}